\documentclass[12pt]{article}

\usepackage{amsmath,amssymb,array,amsfonts}%{{{
\usepackage[dvipdfmx]{graphicx}
\usepackage{comment,color}
\usepackage{ascmac}
\usepackage{cite}

\newcommand{\ba}{\begin{eqnarray}}
\newcommand{\ea}{\end{eqnarray}}

\allowdisplaybreaks

\def\e {{\rm e}}
\def\B {{\rm B}}
\def\T {{\rm T}}
\def\Tr{{\rm Tr}}
\def\dy{{\rm d}y}
\def\dis{\displaystyle}
\def\nt{\notag}

\makeatletter
    
    \@addtoreset{equation}{section}
\makeatother

\begin{document}
\setlength{\baselineskip}{18pt}

\begin{titlepage}%{{{

\begin{flushright}
				OCU-PHYS 447
\end{flushright}
\vspace{1.0cm}
\begin{center}
{\Large\bf 
				Trilinear gauge boson couplings 
				\vspace*{3mm}
				
				in the gauge-Higgs unification
} 
\end{center}
\vspace{25mm}

\begin{center}
{\large
Yuki Adachi 
%\footnote{e-mail : y-adachi@matsue-ct.ac.jp},
and 
Nobuhito Maru$^{*}$
%\footnote{e-mail : nmaru@sci.osaka-cu.ac.jp}
}
\end{center}
\vspace{1cm}
\centerline{{\it
Department of Sciences, Matsue College of Technology,
Matsue 690-8518, Japan.}}

\centerline{{\it
$^{*}$
Department of Mathematics and Physics, Osaka City University, Osaka 558-8585, Japan.
}}
%
%%%%%%%%%%%%%%%%%%%%%%%%%%%%%%% Abstract %%%%%%%%%%%%%%%%%%%%%%%%%%%%%%%%{{{
%
\vspace{2cm}
\centerline{\large\bf Abstract}
\vspace{0.5cm}

We examine  trilinear gauge boson couplings (TGCs) 
in the context of the 
$SU(3)_W\otimes U(1)'$ gauge-Higgs unification scenario.
The TGCs play important roles in the probes of the physics beyond the standard model,  since they are highly restricted by the experiments. 
We discuss mass spectrum of the neutral gauge boson with brane-localized mass terms carefully 
and find that the TGCs and $\rho$ parameter may deviate from standard model predictions. 
Finally we put a constraint from these observables 
and discuss the possible parameter space. %}}}

\end{titlepage}%}}}

\newpage
%%%%%%%%%% begin section  %%%%%%%%%%%%%%%%%%%%%%
\section{Introduction}%{{{

Gauge-Higgs unification (GHU) \cite{GH,HIL} is a scenario that unify the standard model (SM) gauge boson and higgs boson into the 
higher dimensional gauge fields.
It is  one of the attractive ideas that can solve the hierarchy problem without invoking supersymmetry, 
since the higgs boson mass and its potential are  calculable due to the higher dimensional gauge symmetry\cite{HIL}.
These characteristic features  have been studied by explicit diagrammatic calculations 
 and verified in models with various types of compactification 
 at one-loop level %\footnote{For the case of gravity-gauge-Higgs unification, see \cite{HLM}.} 
 \cite{ABQ}
 and at the two-loop level \cite{MY}. 
The finiteness of other physical observables such as $S$ and $T$ parameters \cite{LM}, 
 Higgs couplings to digluons, diphotons \cite{Maru}, muon $g-2$ and the EDM of neutron \cite{ALM} 
 have been investigated. 
 The flavor physics which is a very nontrivial issue in GHU has been studied in \cite{flavorGHU}.

Recent reports on the yukawa couplings in the gauge-Higgs unification scenario 
\cite{GHYukawaWarped,GHYukawaflat,Adachi:2015ova}
show that 
the yukawa couplings become nonlinear functions of vacuum expectation value (VEV) $v$ of Higgs boson
and may deviate from the SM predictions.
In this scenario, higgs fields are a part of the higher dimensional gauge fields 
so that the VEV  becomes periodic in $2/(gR)$
because the yukawa couplings originated from the gauge interactions appear with following Wilson line phase form
\begin{equation}
		W
		=
		P \exp 
		\left[i\frac{g}{2}\oint_{S^1} A_y^{(0)}\dy \right]
		=
		P \exp 
		\left[ig\pi R v \right]
\end{equation}
where $g$ and $R$ stands for the four-dimensional gauge coupling and compactification scale, respectively.
The kink mass for the fermion 
are also required to realize the yukawa couplings for the light fermions,
and then,
the non trivial mixings between the different KK mode appear since the kink mass  breaks translational invariance of the fifth dimension.
Such mixings avoid level crossing in a large VEV,
then the yukawa couplings and the mass spectrum becomes nonlinear functions of $v$.
Namely, the key of this  mechanism is an interplay between the non-vanishing VEV 
and the fermion kink mass. 
They are generic features in the Randall-Sundrum space-time 
\cite{GHYukawaWarped}
and flat space-time
\cite{GHYukawaflat}.

From  this point of view, such deviations may appear not only in the yukawa couplings but also  in gauge boson couplings. 
In fact, 
we consider the $SU(3)_W\otimes U(1)'$ GHU model 
and  find that the trilinear gauge boson couplings (TGCs) and $\rho$ parameter become
nonlinear function of the VEV even at the tree level.
In this model, the $SU(3)_W$ gauge symmetry breaks down to $SU(2)\otimes U(1)$ by the $Z_2$ symmetry,
the SM $Z$ boson is identified as a mixture of remnant $U(1)$  and
extra $U(1)'$ gauge bosons.
These mixing yields the correct weak mixing angle\cite{SSS}.
Since  another combination of gauge boson ($X$) is anomalous,
the brane-localized mass term of the gauge boson appears and becomes massive.
Such brane-localized mass terms break the translational invariance, 
the gauge boson couplings are expected to be the function of VEV
 similar to the yukawa couplings.
Possible deviations  in the gauge couplings are phenomenologically important,
since the TGCs play important roles as the probe of new physics.

This paper is organized as follows. 
In section 2, we introduce our model  
and discuss the equations of motion and the corresponding boundary conditions for the gauge bosons. 
Analytic expression of the $\rho$ parameter and TGCs are derived in section 3. 
Numerical calculations for these parameters  are performed 
and a constraint and a possible parameter space are found. 
Section 4 is devoted to summary. 
In appendix A, the derivation of the equations of motion of the gauge boson 
and its solutions are  described in detail. 

%}}}
%%%%%%%%%% end section  %%%%%%%%%%%%%%%%%%%%%%

%%%%%%%%%% begin section  %%%%%%%%%%%%%%%%%%%%%%
\section{The Model}%{{{
%%%%%%%%%%%%%%%%%%%%%%%%%%%%%%%%

We consider  an
$SU(3)_W\otimes U(1)'$ 
gauge theory in five dimensions compactified on $S^1/Z_2$
where the radius of $S^1$ is $R$.
The strong interaction and fermion sector are omitted since we are interested in the TGCs and $\rho$ parameter of the electroweak sector at tree level. 
The $SU(3)_W$ sector contains the $SU(2)\otimes U(1)$ gauge boson 
and the Higgs doublet corresponds to the coset space ($SU(3)/SU(2)$).
As was mentioned in the introduction, 
the $SU(3)_W$ gauge symmetry is  broken to $SU(2)\otimes U(1)$ by the orbifolding,
but the predicted weak mixing angle $\theta_W$ is too large.
%Furthermore, the hypercharge of the top quark is too small 
%because it is embedded in the four rank totally symmetric tensor representation $\overline{\bf 15}_{-2/3}$
%where the subscript represents the $U(1)'$ charge
%\cite{Cacciapaglia:2005da}.
Furthermore, the higher dimensional representation such as a four rank totally symmetric tensor representation $\overline{\bf 15}_{-2/3}$ is required to realize the yukawa coupling for the top quark
\cite{Cacciapaglia:2005da}
.
However, the hypercharge of the top quark is too small.
These inconsistencies are fixed by introducing the extra $U(1)'$ gauge symmetry.
The $U(1)_Y$ gauge boson in this model is the mixture of the $U(1)$ and $U(1)'$ gauge bosons, 
another linear combination $Z'$ is anomalous
so that the remnant massless gauge bosons are  $SU(2)\otimes U(1)_Y$.

%%%%%%%%%%%%%%%%
\subsection{The Lagrangian}
%%%%%%%%%%%%%%%%

The Lagrangian of the gauge sector consists of the gauge kinetic terms,
gauge fixing term $\mathcal L_{\rm GF}$ and brane-localized mass term $\mathcal L_{\rm B}$.
\begin{equation}
		 \mathcal L_{\rm G}
		 =
		 -\frac12 \Tr F_{MN}F^{MN} - \frac14 B^{MN}B_{MN}
		 + \mathcal L_{\rm GF}
		 +\mathcal L_{\B}
	\label{}
\end{equation}
where the capital letters are understood to be an index of five dimensions $M=0,1,2,3,5$.
The field strength  of $SU(3)_W$ and $U(1)'$ are defined by
\begin{equation}
	F_{MN}=F_{MN}^a T^a 
	=(\partial_M A_N^a -\partial_{N} A^a_{M} +g_5 f^{abc}A_{M}^b A_{N}^c )T^a
	\,,\,
	B_{MN} 
	=\partial_M B_{N} -\partial_{N} B_{M}
	\,,
	\label{}
\end{equation}
where the $f^{abc}$  represents the structure constant of $SU(3)_W$.
The $T^a$ is the generator of the $SU(3)_W$.
%namely, $T^a=\frac{1}{2} \lambda^a$.
The $g_5$ represents the five dimensional gauge coupling 
for the $SU(3)_W$.
The explicit form of $SU(3)_W$ gauge fields are 
\begin{equation}
	A^a T^a
	=
	\frac12
	\begin{pmatrix}
	 A^3 +\frac{2}{\sqrt{6}} A^8 & A^1-iA^2 & A^4-iA^5
	\\
	A^1+iA^2 & - A^3 +\frac{2}{\sqrt{6}} A^8 & A^6-iA^7
	\\
	A^4+iA^5 & A^6+iA^7 & -\frac{4}{\sqrt{6}}A^8
	\end{pmatrix}.
	\label{}
\end{equation}
The gauge-fixing terms are given as follows 
\begin{align}
	\mathcal L_\text{GF} 
	=& 
	-\frac{1}{2\xi}
	\left[\partial^\mu A_\mu^a +\xi(\partial_yA_y^a+2M_Wf^{ab6}A_y^b)  \right]^2
	-\frac{1}{2\xi'}\left[ \partial^\mu A_\mu + \xi' \partial_yA_y\right]^2
	\label{}
\end{align}
where $\xi$ and $\xi'$ stand for gauge fixing parameters
of the $SU(3)_W$ and $U(1)'$, respectively.
The brane-localized gauge boson mass terms reflecting the gauge anomaly  are given by
\begin{equation}
	\mathcal L _\B =\frac{1}{2} M_G^2 \pi R(\delta (y)+\delta (y-\pi R)) Z'^{M}Z'_{M}
\end{equation}
where $M_G$ stands for the brane-localized mass.
The $Z'$ gauge boson, which is a mixture of $A_8$ and $B$, is an  anomalous gauge boson. 

We parameterize these mixings by $\theta$ and $\theta_W$ as 
\begin{equation}
	 \begin{cases}\dis
		Z'&=\cos\theta B -\sin \theta A_8
		\\
		Y&=\cos\theta A_8 +\sin \theta B
	 \end{cases} 
	 ~~~,~~~
	 \begin{cases}\dis
		Z&=\cos\theta_W A_3 - \sin\theta_W Y 
	%	=\frac{1}{\sqrt{1+3\cos^2\theta} }
	%	[A_3-\sqrt3 \cos^2\theta A_8-\sqrt{3}\cos\theta \sin\theta B ]
		\\
		\gamma& = \cos\theta_W Y +\sin\theta_W A_3
	%	=\frac{1}{\sqrt{1+3\cos^2\theta}  }
	%	[\sqrt3 \cos\theta A_3 +\cos\theta A_8 +\sin\theta B ]
	 \end{cases} 
\end{equation}
where the $\theta_W$ represents the weak mixing angle.
To investigate how the neutral gauge bosons mix each other,
we extract the electromagnetic current.
%{color{red}
The down-type quarks are included in the 
%}
${\bf 3} _0=(u,d,d)^{\rm T}$.
\begin{equation}
		 \frac12 g_5 (A_\mu^3\lambda^3+A^8_\mu \lambda^8)
		 \supset
		 \left[\frac23 e_5,-\frac13 e_5 , -\frac13 e_5\right]\gamma_\mu
\end{equation}
where the $e_5$ stands for the five dimensional electromagnetic coupling.
As for the $\overline{\bf 15}_{-2/3}$, 
the right-handed top quark corresponds to the $SU(2)$ singlet,
so we have
\begin{equation}
		 \frac12 \times (-1) \times(-2\frac{\sqrt{3}}{3})\times 4\times  g_5 A^8_\mu 
		 -\frac23 g_5' B_\mu
		 \supset 
		 \frac23 e_5 \gamma_\mu
\end{equation}
where $g_5'$ is the five dimensional gauge coupling for the $U(1)'$.
The first term consists from 
the normalization of the $\lambda^8$ ,
the negative sign which reflects the complex representation,
the eigenvalues for the $U(1)$
and 
the number of the indices of $\overline{\bf 15}_{-2/3}$.
Then these mixings can be read off as 
\begin{equation}
		\tan\theta_W=\sqrt{3} \cos\theta\,,\,
		\cos\theta=\frac{g_5'}{\sqrt{3g_5^2+g_5'^2} } \,.
\end{equation}
The $g_5$ and $g_5'$ stand for the five dimensional gauge couplings of
$SU(3)_W$ and $U(1)'$, respectively.

\subsection{Boundary condition}
We require a periodic boundary condition for the gauge fields along the $y-$direction as 
\begin{equation}
		A_M(y+2\pi R) = A_M(y).
	\label{periodic_BC}
\end{equation}
To break the $SU(3)_W$ gauge symmetry, we furthermore require the $Z_2$ parity at the origin $y=0$ as  
\begin{equation}
	\begin{cases}
		A_\mu(x^\mu,y)= P^\T T^a A_\mu^a (x^\mu,-y) P 
		\,,
		\\
		A_y(x^\mu,y)= -P^\T T^a A_y^a (x^\mu,-y) P\,, 
	\end{cases}
	\label{Z2_BC}
\end{equation}
where $P=\text{diag} (++-)$ for $SU(3)_W$ and $P=1$ for $U(1)'$.

%%%%%%%%%%%%%%%%
\subsection{Mass spectrum and mode functions}
%%%%%%%%%%%%%%%%
In this subsection we discuss the mode functions and its mass spectrum 
which is necessary for calculating TGCs.
There are  two kinds of mixings between the neutral gauge bosons in terms of 
the Higgs VEV $\langle A_y^{6(0)}\rangle=v$ and brane-localized gauge mass terms.
We completely solve these mixings and obtain the mode functions. 
Since the TGCs are defined by the couplings between the charged gauge boson and neutral gauge boson, 
we focus on the zero mode gauge bosons.
Detailed  arguments are included in the appendix \ref{Appendix1}.

The  quadratic terms of the Lagrangian $\mathcal L_{\rm G}$ are 
extracted  as follows.
\begin{align}
		 \mathcal L_{\rm G}
	\supset&
	-\frac14 (\partial_\mu A_\nu-\partial_\nu A_\mu)^2 
	+\frac12(\partial_y A_\mu)(\partial_y A_\mu) 
	-\frac{1}{2\xi'}(\partial^\mu A_\mu)^2
	\nt \\
	&+\frac12 (\partial_\mu A_y)(\partial^\mu A_y) -\frac12 \xi'(\partial_y A_y)^2
	\nt \\
	&-\frac14 (\partial_\mu A_\nu^a-\partial_\nu A_\mu^a)^2 
	+\frac12 (\partial_y A_\mu^a +2M_Wf^{ab6}A_\mu^b)^2
	-\frac{1}{2\xi}(\partial^\mu A^a_\mu)^2
	\nt \\
	&+\frac12 (\partial_\mu A_y^a)(\partial^\mu A_y^a) 
	-\frac12 \xi(\partial_y A_y^a+2M_Wf^{ab6}A_y^b)^2\,.
	\label{}
\end{align}
The mixing terms in the quadratic terms are completely cancelled out by 
choosing suitable gauge-fixing terms.
Hereafter, we choose 
the 't Hooft-Feynman gauge ($\xi=\xi'=1$) 
for simplicity.
We also treat the $U(1)'$ gauge field $B_\mu$ as $A^0_\mu$,
and hence, the equation of motion (EOM) for the gauge fields becomes 
\begin{align}
		 &\left[ \Box \delta^{bc}
	-(\partial_y \delta^{ba}+2M_W f^{ba6})
	(\partial_y \delta^{ac}+2M_W f^{ac6})
 	\right]A ^c 
	\nt\\
	&~~~~~~~~~~~~~~~=
	-\frac12 \pi RM_G^2 [\delta(y) +\delta(y-\pi R) ] 
	\frac{\partial} {\partial A^b} [\cos\theta A^0-\sin\theta A^8] ^2 
\end{align}
where the Lorentz indices are omitted.

By expanding in terms of the mode function, 
the d'Alembertian $\Box$ is replaced with the mass eigenvalue $-m^2$. 
Decomposing into  
charged gauge boson($a=1,2,4,5$) and neutral gauge boson ($a=0,3,7,8$),
we have the following EOMs for the charged gauge boson
\begin{equation}
		 -m^2A
		 =
		 (\partial_y+M_WM_{\rm C})	
		 (\partial_y+M_WM_{\rm C})A	
		 \label{EOM_charged}
\end{equation}
where
\begin{equation}
		 M_{\rm C}=
	\begin{pmatrix}
		0&0&0&-1
		\\
		0&0&+1&0
		\\
		0&-1&0&0
		\\
		+1&0&0&0
	\end{pmatrix}\,,
	\label{EOM_charged1}
\end{equation}
and for the neutral gauge boson
\begin{equation}
				-m^2 A
				=
				( \partial_y +2M_WM_{\rm N})
				( \partial_y +2M_WM_{\rm N})
				A
				+\pi R M_G^2 [\delta(y)+\delta(y-\pi R)  ] U^\dag {\rm diag} (1,0,0,0) U A\,,
		 \label{EOM_neutral}
\end{equation}
where
\begin{equation}
		 M_{\rm N} =
				\begin{pmatrix}
						 0&0&0&0
						 \\
						 0&0 & \frac12 &0
								\\
						 0&-\frac12 & 0 & \frac{\sqrt{3}}{2}
								\\
						0&0&-\frac{ \sqrt{ 3}}{ 2} &0
				\end{pmatrix}\,,\, 
				U=\begin{pmatrix}
					\cos\theta &0&0&-\sin\theta
					\\
					0&1&0&0
					\\
					0&0&1&0
					\\
					\sin\theta&0&0&\cos\theta
		 \end{pmatrix}\,.
\end{equation}
The higgs VEV $v$ is involved in the $M_W=gv$ where 
$g$ is the four dimensional gauge coupling $g=\frac{g_5}{\sqrt{2\pi R}}$.
Solving the above EOM with the boundary conditions
eq(\ref{periodic_BC}) and eq(\ref{Z2_BC}),
we obtain the following mode functions and its mass spectrum.

Let us first discuss the charged gauge boson.
The SM charged gauge boson $W^{\pm}_\mu(x)$  can be read off as
\begin{equation}
		 A^1_\mu(x,y)
		 \supset 
		 \frac{1}{\sqrt{2\pi R}} \frac{W^+_\mu(x)+W^-_\mu(x)}{\sqrt2}
		 ,\,
		 A^2_\mu(x,y)
		 \supset
		 \frac{1}{\sqrt{2\pi R}} \frac{-W^+_\mu(x)+W^-_\mu(x)}{\sqrt2 i}
		 \,.
\end{equation}

As for the neutral gauge boson,
we solve the EOM and extract zero mode similar to the charged gauge boson.
Since the brane mass terms are generated at the cutoff scale,
such as a Grand Unified Theory,
we take the limit $M_G\to \infty$.
Because the EOM is solved by factoring out the VEV  $v$ or $M_W$ 
as shown in the appendix \ref{Appendix1},
we discuss on the $\hat A$ basis 
which are defined by eq. (\ref{hatA})
\begin{align} 
	 \label{mixing0378}
	 \begin{pmatrix}
		A^0\\A^3\\A^7\\A^8	
	 \end{pmatrix}
	 =&
	 \begin{pmatrix}
		\cos\theta\hat A^0+\sin\theta \hat A^8
		\\
		(\frac34 +\frac14\cos2M_Wy)\hat A^3 -\frac12\sin 2M_Wy \hat A^7+
		\frac{\sqrt3}{4}(1-\cos2M_Wy)(\cos\theta\hat A^8 -\sin\theta\hat A^0)   
		\\
		\frac12\sin2M_Wy\hat A^3 + \cos 2M_Wy\hat A^7-\frac{\sqrt3}{2}\sin2M_Wy(\cos\theta\hat A^8-\sin\theta \hat A^0)  
		\\
		\frac{\sqrt3}{4}(1-\cos2M_Wy)\hat A^3+\frac{\sqrt3}{2}\sin2M_Wy\hat A^7
		+(\frac{1}{4}+\frac{3}{4}\cos2M_Wy)(\cos\theta \hat A^8-\sin\theta\hat A^0)  
	 \end{pmatrix}\,,
\end{align}
where $\cos\theta=\frac{\sin\theta_W}{\sqrt3\cos\theta_W},\sin\theta=\frac{\sqrt{4\cos^2\theta_W-1}}{\sqrt3\cos\theta_W}$.
In this basis,  the SM photon $\gamma$ and $Z$ boson are extracted as  
\begin{equation}
		 \begin{cases}
					\hat A^3_\mu (x^\mu,y)
					\supset 
					\sin\theta_W\gamma_\mu (x^\mu) f^0_\gamma (y)\,, 
					\\
					\hat A^8_\mu(x^\mu,y)
					\supset 
					\cos\theta_W \gamma_\mu(x^\mu) f^0_\gamma (y)\,,
		 \end{cases}
\end{equation}
and 
\begin{equation}
		 \begin{cases}
					\hat A^0_\mu(x^\mu,y)
					\supset 
					 \sqrt{\frac{4\cos^2\theta_W-1}{4\cos^2\theta_W-\sin^2\hat M_W}}\sin \hat M_W Z_\mu(x^\mu)f^0_Z(y)
					\,,
					\\
					\hat A^3_\mu(x^\mu,y)
					\supset 
					\cos\theta_W Z_\mu(x^\mu)f^3_Z(y)
					\,,
					\\
					\hat A^7_\mu(x^\mu,y)
					\supset  
					-\frac{2 \cos\theta_W\cos\hat M_W}{\sqrt{4\cos^2\theta_W-\sin^2\hat M_W}} Z_\mu(x^\mu)f^7_Z(y)
					\,,\\
					\hat A^8_\mu(x^\mu,y)
					\supset 
					-\sin\theta_W Z_\mu(x^\mu)f^8_Z(y)
					\,,
		 \end{cases}
\end{equation}
where the dimensionless $W$ boson mass parameter is introduced $\hat M_W=\pi R M_W$. 
The mode  functions
are obtained as follows.
\begin{equation}
 \begin{cases}
			f^0(y)=
	-\frac{1}{\sqrt{\pi R-\frac{1}{2m}\sin2\pi Rm}}\sin m|y|
	,\\
	f^3(y)=f^8(y)=
	\frac{1}{\sqrt{\pi R+\frac{1}{2m}\sin2\pi Rm}}\cos my
	,\\
	f^7(y)=\frac{1}{\sqrt{\pi R-\frac{1}{2m}\sin2\pi Rm}}\sin my
	\,.
 \end{cases}
\end{equation}
The subscripts $\gamma$ and $Z$ are understood to substitute the corresponding mass eigenvalues.
The mass spectrum is given by the solutions of 
\begin{align}
		 \label{gammaZbosonmass}
		 \sin^2\hat m_\gamma =& 0\,,\, 
		 \tan\hat m_Z =
		 \frac{\sqrt{4\cos^2\theta_W-\sin^2\hat M_W}}{2\cos^2\theta_W-\sin^2\hat M_W}\sin\hat M_W
\,.
\end{align}
The derived mass eigenvalue $m_Z$ is found 
{\it i.e.},
$m_Z=M_Z(v) +\frac{n}{R}$,
so that 
the  $Z$ boson mass $M_Z(v)$ corresponds to the minimal values of $m_Z$.

%}}}
%%%%%%%%%% end section  %%%%%%%%%%%%%%%%%%%%%%

%%%%%%%%%% begin section  %%%%%%%%%%%%%%%%%%%%%%
\section{$\rho$ parameter and Trilinear gauge boson couplings}%{{{

We now focus on the $\rho$ parameter 
and TGCs.
As was mentioned earlier, 
these couplings or the parameter may deviate from the SM predictions 
even at the tree level
because of the nonlinearity of higgs VEV. 
Naively, this fact is very phenomenologically dangerous 
since these parameters have been precisely measured by experiments 
and the severe constraints for them are provided. 
Therefore, we should investigate whether our model satisfies these constraints. 
After the analytic expressions of the $\rho$ parameter and TGCs
are derived,
we perform the numerical study.

\subsection{$\rho$ parameter}%{{{
The $\rho$ parameter is defined by the ratio among  the $W$ boson mass, 
$Z$ boson mass and weak mixing angle:
\begin{equation}
		\rho
		=
		\frac{M_W}{\cos\theta_W M_Z(v)}
		\,.
\end{equation}
$\rho=1$ at the tree level in the SM since the $Z$  boson mass $M_Z(v)$ is given by $M_W/\cos\theta_W$ at the tree level.
However, 
the $\rho$ parameter in our model is dependent on $v$ 
because the $Z$ boson mass is nonlinear function of $v$, 
{\it i.e.} $m_Z=M_Z(v)$.
It is determined by the  relation (\ref{gammaZbosonmass}),
the $\rho$ parameter in our model is defined as 
%Although the $\rho$ parameter is constant in the SM, 
%since the Z boson mass $M_Z$ depends on the VEV $v$,
%the $\rho$ parameter also depends on it.
%An explicit form is given by
\begin{equation}
		\rho
		=
		\frac{1}{\cos\theta_W }
		\frac{\hat M_W}{\tan^{-1}\left[\frac{\sqrt{4\cos^2\theta_W-\sin^2\hat M_W}}{2\cos^2\theta_W-\sin^2\hat M_W}\sin \hat M_W\right]}
		\,.
\end{equation}
Note that the arctangent in the denominator 
stands for the minimal values.
%We can easily show that the  $\rho$ parameter becomes $1$ while the {\color{red} $\hat M_W\ll 1$}.
The $\rho$ parameter in our model agrees with the SM one 
 in the linear limit of $v$.
Once the nonlinearity of $v$ is taken into account,
it deviates from 1.

It is notable that the $\rho$ parameter reduces to $1$ 
in the limit $\cos^2\theta_W\to 1/4$ , namely, $\theta\to 0$.
It is easy to  understand
since the the brane-localized mass term couples to the $U(1)'$ gauge fields only. 
Therefore, 
the translational invariance 
for the $SU(3)_W$ gauge fields
is kept in this limit.
Then, such deviation of the $\rho$ parameter vanishes.

%}}}

\subsection{Trilinear gauge boson couplings}%{{{
In this subsection, we discuss  the TGCs 
which are highly restricted from the several experiments.
They are parameterized in the following form
\cite{Lopez-Osorio:2013xka}
\begin{equation}
	 \mathcal L_\text{TGC} =
	 -ig_V 
	 \left[
	 g_1^V(W_{\mu\nu}^+ W^{-\mu}V^\nu -W_{\mu\nu}^- W^{+\mu}V^\nu)
	 +\kappa_V W_\mu^+ W_\nu^-V^{\mu\nu} 
	 +\frac{\lambda_V}{M_W^2}W_{\mu\nu}^+ W^{-\nu\rho}V_\rho^\mu 
\right]
\end{equation}
where $W^\pm _{\mu\nu}=\partial_\mu W^\pm_\nu-\partial_\nu W^\pm_\mu$
and $V_{\mu\nu}=\partial_\mu V_\nu-\partial_\nu V_\mu$.
The $V$ represents the neutral gauge boson 
{\it e.g.}, $\gamma$
and $Z$ boson.
The coupling $g_V$ corresponds to $g_\gamma=\sin\theta_W g$ and $g_Z=\cos\theta_W g$ in the SM.
They are restricted  as 
\begin{equation}
		 \label{TGC_exp}
	 -0.057 < \Delta \kappa_\gamma < 0.154
%	 \,,\,
%	 -0.015< \lambda < 0.028
	 \,,\,
	 -0.008 < \Delta g_1^Z < 0.054\,.
\end{equation}
by the experiments 
\cite{Abazov:2012ze}.
The $\Delta \kappa$ and $\Delta g_1^Z$ defined by $\kappa -1$ and $g_1^Z -1$, respectively. 

The TGCs in this model is given by extracting the terms which couples to the  charged gauge boson $W^\pm_\mu(x^\mu)$ of the SM  
from the Lagrangian as 
\begin{align}
		 \label{TGC}
	 \mathcal L_\text{TGC}
	 =&
	 \frac{g_5}{\sqrt{2\pi R}}
	 \int_{-\pi R} ^{\pi R}\dy
	 \left[
				iW_{\mu\nu}^+(x^\mu) W^{-\mu}(x^\mu)A^{3\nu}(x^\mu,y) + {\rm h.c.} 
				+i f_{\mu\nu}^3(x^\mu,y) W^{+\mu}(x^\mu)W^{-\nu}(x^\mu)
	 \right]
\end{align}
where $f^3_{\mu\nu}=\partial_\mu A^3_\nu-\partial_\nu A^3_\mu$.
Since the SM charged gauge boson $W^{\pm}_\mu$ only couple to $A^3_\mu$,
the TGCs in this model are given by 
 substituting following explicit form 
\begin{align}
		 \label{A3}
		 A^3_\mu(x^\mu,y)
		 \supset&
		 \sin\theta_W f_\gamma^3(y)\gamma_\mu(x^\mu) 
		 -
		 \frac
		 {(4\cos^2\theta_W-1)\sin\hat M_W[1-\cos(2M_Wy)]}
		 {4 \cos\theta_W\sqrt{4\cos^2\theta_W-\sin^2\hat M_W}}
		 f^0_Z(y)Z_\mu(x^\mu)
		 \nt\\
		 &
		 +\frac
		 {2\cos^2\theta_W+1+(2\cos^2\theta_W-1)\cos (2M_Wy)}
		 {4\cos\theta_W} 
		 f_Z^3(y) Z_\mu(x^\mu)
		 \nt\\
		 &+
		 \frac
		 {\cos\theta_W\cos\hat M_W\sin (2M_Wy) }
		 {\sqrt{4\cos^2\theta_W-\sin^2\hat M_W}}
		 f^7_Z(y)Z_\mu(x^\mu)
		 \,.
\end{align}

We find the TGC for the photon  as follows:
		\begin{align}
		 \mathcal L_{\rm TGC}
		 \supset&
		 \int_{-\pi R}^{\pi R} \dy 
		 \left[ig_5W_{\mu\nu}^+ W^{-\mu}\sin\theta_W\gamma^\nu\frac{1}{(2\pi R)^{3/2}}
		 +{\rm h.c.}
		+ig_5\gamma_{\mu\nu}W^{+\mu}W^{-\nu}\frac{1}{(2\pi R)^{3/2}}\right]
		\nt \\
		=&
		i\frac{g_5\sin\theta_W}{\sqrt{2\pi R}}W_{\mu\nu}^+W^{-\mu}\gamma^\nu
		+{\rm h.c.}
		+i\frac{g_5\sin\theta_W}{\sqrt{2\pi R}}\gamma_{\mu\nu}W^{+\mu}W^{-\nu}
		\end{align}
		Thus we have
		\begin{equation}
			g_\gamma g_1^\gamma = g_\gamma \kappa_\gamma 
			= 
			\frac{g_5\sin\theta_W}{\sqrt{2\pi R}}.
		\end{equation}

The TGCs for the $Z$ boson are obtained similarly. 
Note the coefficients $\kappa_Z$ and $g_Z$ are same  
because these deviations are originate from  the mode function of $Z$ boson.
An explicit form is given as follows.
				\begin{align}
				 g_Zg_1^{Z}=&g_Z\kappa_{Z}
				 \nt \\
				 =&
%				 \frac{g}{2\pi R}
%				 \int_{-\pi R}^{\pi R}\dy
%				 \left[
%					\frac{2\cos^2\theta_W+1+(2\cos^2\theta_W-1)\cos2M_Wy}{4\sqrt{2}\cos\theta_W}f_Z^3
%					\right.
%					\nt \\
%				&
%				 \left. 
%				 +\frac{4\cos^2\theta_W\cos\hat M_W\sin2M_Wyf_Z^7
%				 -(4\cos^2\theta_W-1)\sin\hat M_W(1-\cos2M_Wy)f_Z^0}
%				 {4\sqrt2 \cos\theta_W\sqrt{4\cos^2\theta_W-\sin^2\hat M_W}}
%				\right]
%				\nt \\
%				=&
%				\frac{g}{2\pi R}
%				\frac{2}{4\sqrt{2}\cos\theta_W}
%				 \int_{0}^{\pi R}\dy
%				 \left[
%				(2\cos^2\theta_W+1)f_Z^3+(2\cos^2\theta_W-1)\cos2M_Wyf_Z^3
%				\right.
%				\\
%				&
%				 \left. 
%				 +\frac{4\cos^2\theta_W\cos\hat M_W\sin2M_Wyf_Z^7
%				 +(4\cos^2\theta_W-1)\sin\hat M_W(1-\cos2M_Wy)f_Z^7}
%				 {\sqrt{4\cos^2\theta_W-\sin^2\hat M_W}}
%				\right]
%				\\
%				=&
				\frac{g}{\sqrt{\pi R}}\frac{1}{4\cos\theta_W}
				\nt \\
				&
				\Bigg[
				 \frac{\sin\hat m_Z\sqrt{\hat m_Z}}{\sqrt{2\hat m_Z+\sin2\hat m_Z}}
				  \frac{2\cos^2\theta_W+1}{\hat m_Z}
				 \nt \\
				 &
				 +\frac{2\cos^2\theta_W-1}{\sqrt{2\hat m_Z+\sin2\hat m_Z}}
				 \frac{\sqrt{\hat m_Z}}{2}
				 \left\{\frac{\sin(2\hat M_W+\hat m_Z)}{2\hat M_W+\hat m_Z}
				+\frac{\sin(2\hat M_W-\hat m_Z)}{2\hat M_W-\hat m_Z}
				 \right\}
				 \nt \\
				 &
				 +\frac{2\cos^2\theta_W\sqrt{\hat m_Z}}{\sqrt{2\hat m_Z-\sin2\hat m_Z}}
				 \frac{\cos\hat M_W}{\sqrt{4\cos^2\theta_W-\sin^2\hat M_W}}
				 \left\{\frac{\sin(2\hat M_W-\hat m_Z)}{2\hat M_W-\hat m_Z}
				-\frac{\sin(2\hat M_W+\hat m_Z)}{2\hat M_W+\hat m_Z}
				 \right\}
				 \nt \\
				 &
				 +\frac{(4\cos^2\theta_W-1)\sqrt{\hat m_Z}}{\sqrt{2\hat m_Z-\sin2\hat m_Z}}
				 \frac{\sin\hat M_W}{\sqrt{4\cos^2\theta_W-\sin^2\hat M_W}}
				 \nt \\
				 &~~\times
				 \left\{\frac{1-\cos\hat m_Z}{\hat m_Z}
				 -\frac12 \left(\frac{1-\cos(2\hat M_W+\hat m_Z)}{2\hat M_W+\hat m_Z}-\frac{1-\cos(2\hat M_W-\hat m_Z)}{2\hat M_W-\hat m_Z}\right)\right\}
				\Bigg]\,.
		\end{align}
%		where
%		we use
%		\begin{equation}
%			\begin{cases}
%				 \dis
%				 \int_0^{\pi R}\dy f_Z^3 
%				 =
%				 \frac{1}{\sqrt{\pi R +\frac{1}{2m_Z}\sin2\hat m_Z}}
%				 \int_0^{\pi R}\dy  \cos m_Z y
%				 =
%				 \frac{\sin\hat m_Z}{\sqrt{2\hat m_Z+\sin2\hat m_Z}}
%				 \frac{\sqrt2}{\sqrt{m_Z}}
%				 \\
%				 \dis
%				 \int_0^{\pi R}\dy \cos 2M_Wy f_Z^3 
%				 =
%				 \frac{1}{\sqrt{2\hat m_Z+\sin2\hat m_Z}}
%				 \frac{\sqrt{m_Z}}{\sqrt2}
%				 \left[\frac{\sin(2\hat M_W+\hat m_Z)}{2 M_W+ m_Z}
%				+\frac{\sin(2\hat M_W-\hat m_Z)}{2 M_W- m_Z}
%					 \right]
%				 \\
%				 \dis
%				 \int_0^{\pi R}\dy \sin 2M_Wy f_Z^7 
%				 =
%				 \frac{1}{\sqrt{2\hat m_Z-\sin2\hat m_Z}}
%				 \frac{\sqrt{m_Z}}{\sqrt2}
%				 \left[\frac{\sin(2\hat M_W-\hat m_Z)}{2 M_W- m_Z}
%				-\frac{\sin(2\hat M_W+\hat m_Z)}{2 M_W+ m_Z}
%				 \right]
%				 \\
%				 \dis
%				 \int_0^{\pi R}\dy (1-\cos 2M_Wy) f_Z^7 
%				 =
%				 \frac{\sqrt{2m_Z}}{\sqrt{2\hat m_Z-\sin2\hat m_Z}}
%				 \left[\frac{1-\cos\hat m_Z}{m_Z}
%				+\frac12
%				\left(\frac{1-\cos(2\hat M_W-\hat m_Z)}{2 M_W- m_Z}
%				-\frac{1-\cos(2\hat M_W+\hat m_Z)}{2 M_W+ m_Z}
%				\right)
%				 \right]
%			\end{cases}
%		\end{equation} 
The above result reduce to  the SM prediction $g\cos\theta_W$ 
if we take the limit where the nonlinearity of $v$ can be neglected.

%}}}

\subsection{Numerical study}%{{{
\begin{figure}[h]
\center
\def\FIGSIZE{0.35}
\includegraphics[scale=\FIGSIZE]{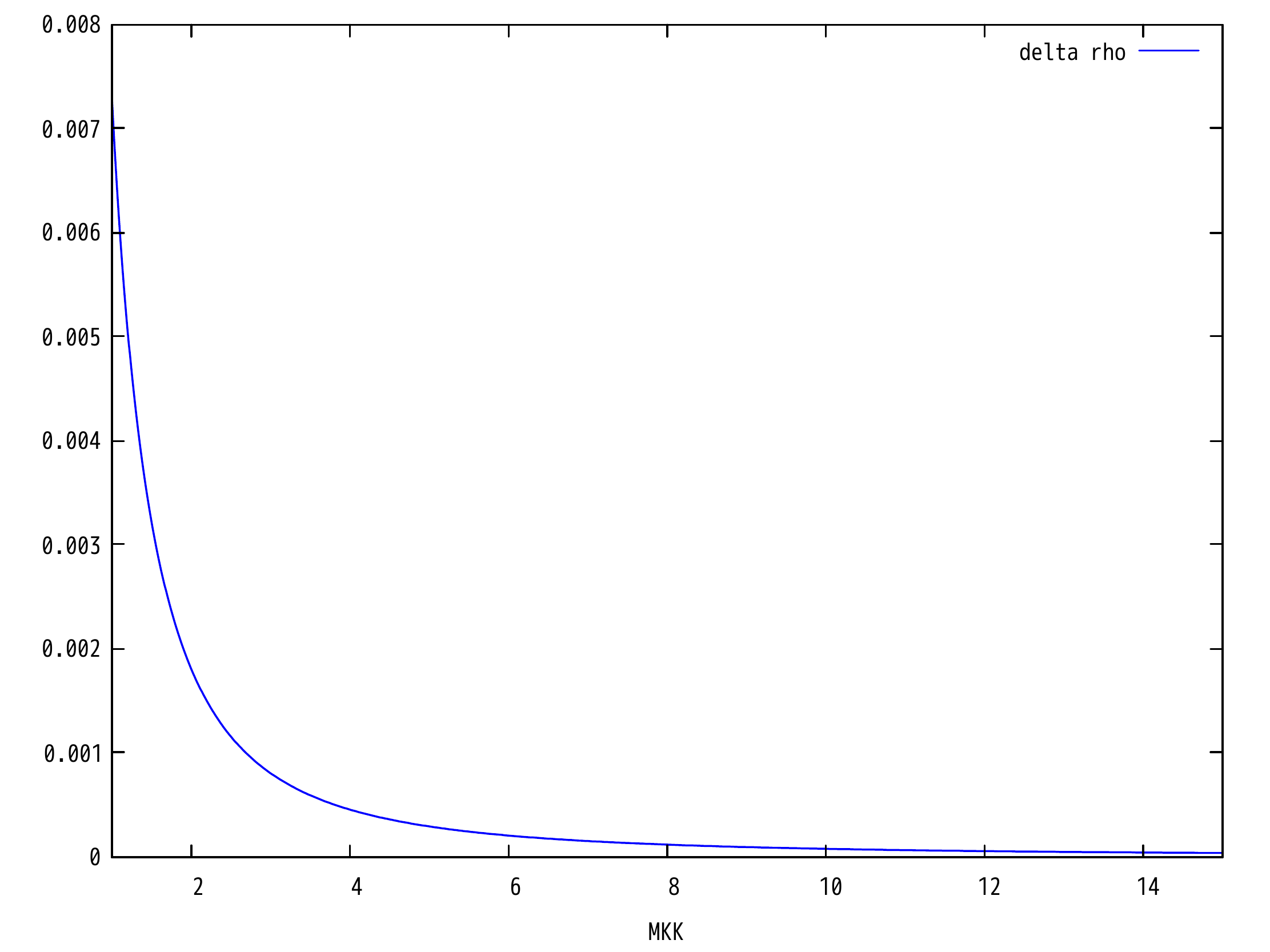}
\includegraphics[scale=\FIGSIZE]{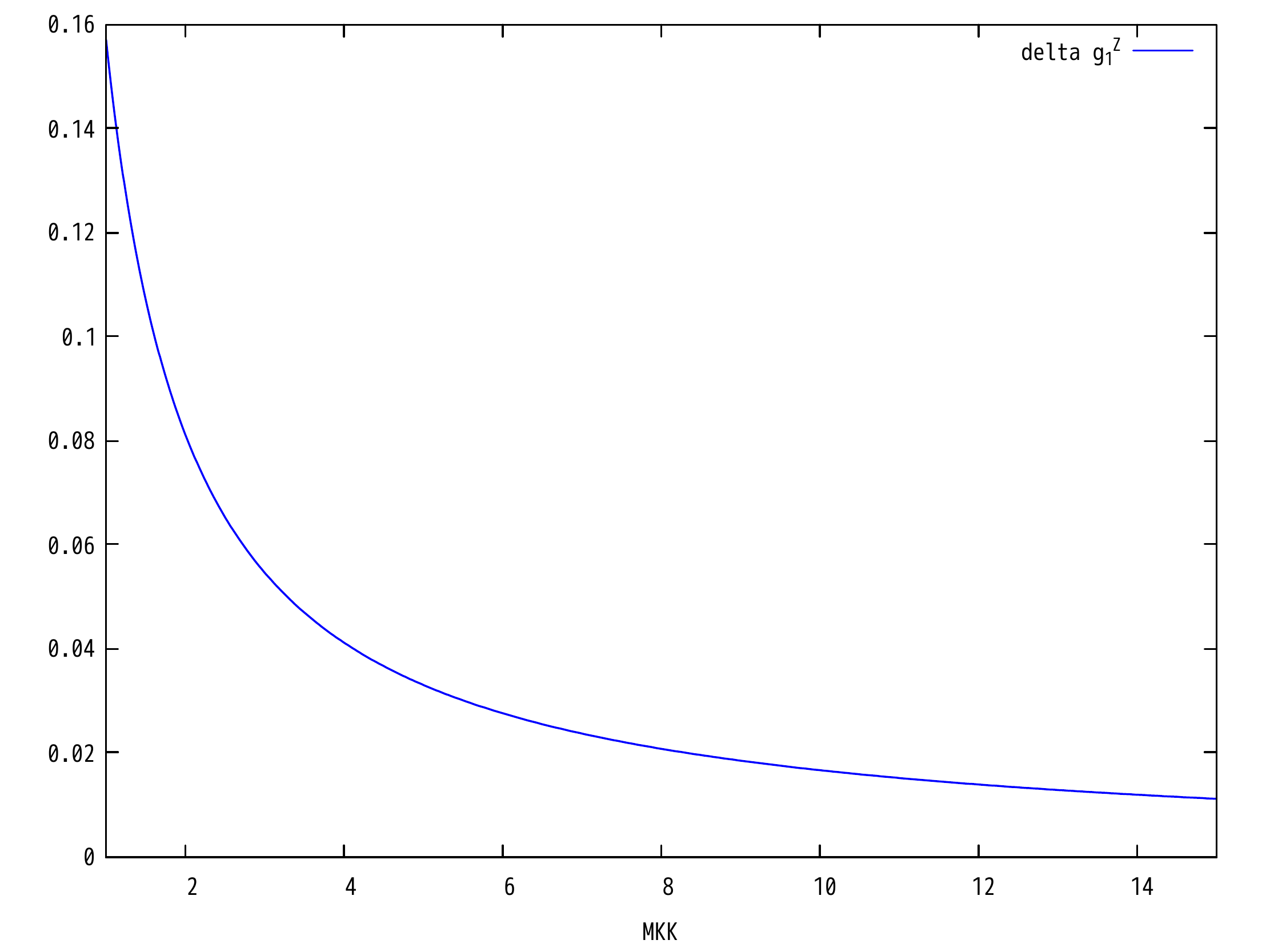}

\caption{The deviation of $\rho$ parameter (left) and $\Delta g_1^Z=\Delta \kappa_Z$ (right) as a function of KK scale are plotted.}
\label{rho_and_g1z}
\end{figure}

In this subsection, we perform a numerical analysis on the TGCs and $\rho$ parameter.
Since the $WW\gamma$ vertex is same as that of  the SM,
we focus on the $WWZ$ coupling. 
The deviation of $\rho$ parameter and the TGCs for $WWZ$ coupling defined by 
\begin{equation}
\Delta \rho = \frac{M_W}{\cos\theta_W M_Z} -1\,,\,\,
\Delta g_1^Z = g_1^Z -1 \,,
\end{equation}
which are depicted in figure \ref{rho_and_g1z}.
Since the $\rho$ parameter is deviated from $1$ even at the tree level,
we hence require that the $\Delta \rho$ is smaller than the contributions from  
the radiative corrections in the SM,
namely,
\begin{equation}
		\Delta \rho =\rho -1 < 0.001\,. 
\end{equation}
From this,
we obtain the lower bound for the compactification scale as
\begin{equation}
		 R^{-1}\geq 3{\rm TeV}\,.		 
\end{equation}
From the constraints of the TGCs of the $WWZ$ coupling eq(\ref{TGC_exp}),
we find
\begin{equation}
		 R^{-1}\geq 3{\rm TeV}\,.		 
\end{equation}

Severer constraints of the TGCs are 
obtained by combining Higgs production data at LHC %\cite{Corbett:2013pja},
$-0.002\leq \Delta g_1^Z\leq 0.026$ \cite{Corbett:2013pja},
we obtain in that case
\begin{equation}
		 R^{-1} \geq 6.3{\rm TeV} \,.
\end{equation}

Finally, we would like to comment on one-loop contributions of nonzero KK modes to TGC. 
%According to the operator analysis, 
These one-loop contributions in our case are also expected to be suppressed very much
%by a loop factor and the square of the ratio between the weak scale and the compactification scale $m_W/M_c$, 
comparing to the SM ones at tree level as shown in \cite{Lopez-Osorio:2013xka}, 
where the one-loop contributions of nonzero KK modes to TGC have been calculated in the universal extra dimensional model.  
Although an issue of quantum corrections to TGC is very interesting and important, 
the calculations are more involved and a very careful analysis is required. 
In particular, KK fermion contributions are model dependent.  
The issue is therefore beyond the scope of this paper and left for a future work. 

%}}}

%}}}
%%%%%%%%%% end section  %%%%%%%%%%%%%%%%%%%%%%

%%%%%%%%%% begin section  %%%%%%%%%%%%%%%%%%%%%%
\section{Summary}%{{{
In this paper, 
we study the $\rho$ parameter and TGCs in the gauge-Higgs unification scenario.
Although they are constants in the SM,  
these couplings or parameter in this model may become  nonlinear functions of  VEV $v$.
It is due to the fact that the translational invariance along with the fifth dimension of this theory 
is broken  down by the brane-localized mass term.
In fact, 
we have derived  the analytic expressions for the $\rho$ parameter and TGCs 
by use of the exact mode functions and its mass eigenvalues
which are given by solving EOM of the neutral gauge bosons.
It indicates  that they are  the function of the VEV $v$.  
We furthermore have verified that they reduce to the SM predictions
in the limit where the nonlinearity of $v$ can be neglected.  
It is quite natural since the VEV $v$  in this scenario is embedded in the Wilson line phase
and it  becomes unit matrix in that limit. 

These deviations are significant in the phenomenological point of view,
because the $\rho$ parameter and TGCs are precisely measured by experiments.
We have performed the numerical study and 
obtained the lower bound of the compactification scale
$R^{-1}> 3{\rm TeV}$.
A severer constraint $R^{-1}> 6.3{\rm TeV}$ is obtained by combining Higgs production data at LHC.
We hope that this result will provide useful information for new physics search 
at LHC Run 2 or ILC in a future. 
 
%We also notify that the $\Delta g_1(\Delta \kappa_{Z})$ is always positive.
%The feature may distinguish the other extra dimensional model  
%since the $\Delta \kappa _Z$ is negative in 
%universal extra dimensional model.

%}}}
%%%%%%%%%% end section  %%%%%%%%%%%%%%%%%%%%%%

%%% Acknowledgments %%%%%%%%%%%%%%%%%%%%%%%%%%%%{{{
%\section*{Acknowledgments}
%The work of N.M. is supported in part by the Grant-in-Aid 
% for Scientific Research from the Ministry of Education, 
% Science and Culture, Japan No. 24540283. 
%}}}

%%%%%
\appendix
%%%%%

%%%%%%%%%% begin section  %%%%%%%%%%%%%%%%%%%%%%
\section{Derivation of mode functions and its KK mass spectrum}%{{{
\label{Appendix1}
In this appendix, we derive the KK mass spectrum of the neutral gauge boson.
As pointed in the main text, there are two kinds of mixings which arise from the brane localized gauge boson mass term and the VEV of the higgs $A_y^6$.  
We completely solve these mixings by 
factoring out the  VEV from mode equations. 
The mass spectrums are determined by  the boundary conditions on the 
mode functions and its derivatives.

\subsection{charged gauge boson} %{{{

The charged gauge boson $W^{\pm}_\mu$ in our model corresponds to the 
$A^1_\mu,A^2_\mu,A^4_\mu,A^5_\mu$ of the $SU(3)_W$.
Their EOM are already derived 
as eq({\ref{EOM_charged}}) and eq({\ref{EOM_charged1}}) 
in the main text.
\begin{equation}
		 -m^2A
		 =
		 (\partial_y+M_WM_{\rm C})	
		 (\partial_y+M_WM_{\rm C})A	
		 \label{EOM_charged}
\end{equation}
where
\begin{equation}
		 M_{\rm C}=
	\begin{pmatrix}
		0&0&0&-1
		\\
		0&0&+1&0
		\\
		0&-1&0&0
		\\
		+1&0&0&0
	\end{pmatrix}\,.
	\label{EOM_charged1}
\end{equation}
Note that we adopt the matrix form
$A=(A^1_\mu,A^2_\mu,A^4_\mu,A^5_\mu)^{\rm T}$.
Let us first eliminate the $M_W$ from the EOM ({\ref{EOM_charged}})
by  defining
$A=\e^{-M_W M_{\rm C} y}{\tilde A}$,
then the EOM becomes
\begin{equation}
				-\partial_y^2 \tilde A =m^2\tilde A  
				\label{EOM_charged_elliminate}
\end{equation}

We require the  {\it $Z_2$ conditions at the origin} and 
{\it periodicity } on the gauge bosons.
The $Z_2$ BCs are the same for both $A$ and $\tilde A$
because of the phase matrix $\exp \left[-M_WM_{\rm C} y\right]$   becomes 
unit matrix at the origin, 
the $Z_2$ condition at the origin 
$\tilde A(x^\mu,y) ={\rm diag}(+,+,-,-)  \tilde A(x^\mu,-y)$.
From the $Z_2$ condition,
the eq (\ref{EOM_charged_elliminate}) is solved as 
\begin{equation}
	\begin{cases}
					\dis
					A^1_\mu(x^\mu,y)
					=
					\sum_n 
					\left[
					\cos M_Wy\cos my \tilde A^{1(n)}_\mu(x^\mu)
					+\sin M_Wy\sin my \tilde A^{5(n)}_\mu(x^\mu)
	\right]
					\\
					\dis
					A^2_\mu(x^\mu,y)
					=
					\sum_n
					\left[
					\cos M_Wy\cos my \tilde A^{2(n)}_\mu(x^\mu)
					-\sin M_Wy\sin my \tilde A^{4(n)}_\mu(x^\mu)
	\right]
					\\
					\dis
					A^4_\mu(x^\mu,y)
					=
					\sum_n
					\left[
					\cos M_Wy\sin my \tilde A^{4(n)}_\mu(x^\mu)
					+\sin M_Wy\cos my \tilde A^{2(n)}_\mu(x^\mu)
	\right]
					\\
					\dis
					A^5_\mu(x^\mu,y)
					=
					\sum_n
					\left[
					\cos M_Wy\sin my \tilde A^{5(n)}_\mu(x^\mu)
					-\sin M_Wy\cos my \tilde A^{1(n)}_\mu(x^\mu)
	\right]
	\end{cases}
\end{equation}
where 
$\e^{-M_W M_{\rm C} y}=\cos M_W y -M_{\rm C}\sin M_W y$
is used.

From the periodicity at $y=\pi R$, 
$A(x^\mu,\pi R) = A(x^\mu,-\pi R)$,
we have
\begin{equation}
				\label{chargedmasscondition1}
				\begin{cases}
								\cos \hat M_W \sin\hat m \tilde A^{4(n)}
								+\sin\hat M_W\cos\hat m \tilde A^{2(n)}
								=0
								\\
								\cos \hat M_W \sin\hat m \tilde A^{5(n)}
								-\sin\hat M_W\cos\hat m \tilde A^{1(n)}
								=0
				\end{cases}
\end{equation}
where the $\hat m$  describes $\pi R m$.
To satisfy the EOM at $y=\pi R$,  
we impose a conditions 
\begin{equation}
				0=
				\lim_{\epsilon\to 0}
				\int_{\pi R-\epsilon}^{\pi R+\epsilon} \dy 
				\left[m^2 +
								(\partial_y+M_WM_{\rm C})	
								(\partial_y+M_WM_{\rm C})
				\right]A\,.
\end{equation}
Since the gauge boson fields $A$ is continuous at $y=\pi R$,
the above condition becomes 
\begin{equation}
				0=\left[\e^{-M_WM_{\rm C}y}\partial _y \tilde A\right]^{\pi R}_{-\pi R}	
				\,
\end{equation}
in the matrix form, or 
\begin{equation}
				\label{chargedmasscondition2}
				\begin{cases}
								0=
								-\cos \hat M_W \sin\hat m \tilde A^{1(n)}	
								-\sin \hat M_W\cos \hat m \tilde A^{5(n)}
								\,,
								\\
								0=
								-\cos \hat M_W \sin\hat m \tilde A^{2(n)}	
								+\sin \hat M_W\cos \hat m \tilde A^{4(n)}
								\,.
				\end{cases}
\end{equation}
These conditions (\ref{chargedmasscondition1})
and (\ref{chargedmasscondition2})
determine the mass spectrum and its eigenstate.
They are summarized in the following form.
\begin{equation}
		 \label{massmatrix_chargedboson}
				0=\begin{bmatrix}
								-\tan \hat m 
								&0
								&0
								&\tan \hat M_W 
								\\
								0
								&\tan \hat m
								&\tan \hat M_W 
								&0
								\\
								-\tan\hat M_W&0&0&\tan\hat m
								\\
								0&\tan\hat M_W & \tan \hat m &0
				\end{bmatrix}
				\begin{pmatrix}
								\tilde A^{1(n)}	\\
								\tilde A^{2(n)}	\\
								\tilde A^{4(n)}	\\
								\tilde A^{5(n)}	
				\end{pmatrix}
				\,.
\end{equation}
The condition that determines the mass spectrum 
is equivalent to 
that the eq(\ref{massmatrix_chargedboson}) has nontrivial solutions,
namely,
the determinant of the matrix in the  eq(\ref{massmatrix_chargedboson}) should be vanished.
This gives two types of spectrum as 
\begin{equation}
		% 0= (\tan\hat m -\tan \hat M_W)^2(\tan\hat m +\tan \hat M_W)^2 
		% \Rightarrow 
		 \tan \hat m=\pm \tan \hat M_W\,.
\end{equation}
The charged gauge boson in the SM  
corresponds to the zero mode of KK modes, {\it i.e.} $m=\pm M_W$,
the $A^1_\mu$ and $A^2_\mu$ is constant with respect to the fifth dimension.
Thus we have 
\begin{equation}
%	\begin{cases}
%		\dis
		 A^1_\mu(x^\mu,y)
		\supset
%		\propto \cos\frac{ny}{R}\tilde A^{1(n)}
%		\Rightarrow 
		\frac{1}{\sqrt{2\pi R}}\frac{W^+_\mu(x^\mu)+W^-_\mu(x^\mu)}{\sqrt2}
%		+\frac{\cos \frac{ny}{R}-\delta_{n0}}{\sqrt{\pi R}}
%		\frac{Y^{+(n)}+Y^{-(n)}}{\sqrt2}
%		\\
		,
		\dis
		A^2_\mu(x^\mu,y)
		\supset
%		\propto \cos\frac{ny}{R} \tilde A^{2(n)}
%		\Rightarrow 
		\frac{1}{\sqrt{2\pi R}}\frac{W^-_\mu(x^\mu)-W^+_\mu(x^\mu)}{\sqrt2 i}.
%		+\frac{\cos \frac{ny}{R}-\delta_{n0}}{\sqrt{\pi R}}
%		\frac{Y^{-(n)}-Y^{+(n)}}{\sqrt2 i}
%		\\
%		\dis
%		A^4\propto \sin\frac{ny}{R} \tilde A^{2(n)}
%		\Rightarrow 
%		\frac{\sin \frac{ny}{R}}{\sqrt{\pi R}}
%		\frac{Y^{-(n)}-Y^{+(n)}}{\sqrt2 i}
%		\\
%		\dis
%		A^5\propto -\sin\frac{ny}{R} \tilde A^{1(n)}
%		\Rightarrow 
%		-\frac{\sin \frac{ny}{R}}{\sqrt{\pi R}}
%		\frac{Y^{+(n)}+Y^{-(n)}}{\sqrt2}
%		\end{cases}
\end{equation}
Note that the factor $1/\sqrt{2\pi R}$ 
comes from the normalization.

%}}}

\subsection{neutral gauge boson}%{{{
In this subsection, we focus on  the neutral gauge boson.
As we mentioned in the introduction,
the $U(1)'$ gauge boson and neutral sector in the $SU(3)_W$ are mixed  
by the boundary term.
Similar to the main text,
we treat the index of $U(1)'$ gauge boson as $a=0$.
Adopting the vector notation
$A=(A^0_\mu,A^3_\mu,A^7_\mu,A^8_\mu)^{\rm T}$,
the EOM for the neutral gauge boson are given by 
\begin{equation}
				-m^2 A
				=
				( \partial_y +2M_WM_{\rm N})
				( \partial_y +2M_WM_{\rm N})
				A
				+\pi RM_G^2 [\delta(y) +\delta(y-\pi R) ] 
				U^\dag {\rm diag} (1,0,0,0) U A
\end{equation}
where the $M_G$ stands for the brane-localized mass term for the anomalous gauge boson.
The matrices $M_\text{N}$ and $U$ are defined by
\begin{equation}
				M_{\rm N} =
				\begin{pmatrix}
						 0&0&0&0
						 \\
						 0&0 & \frac12 &0
								\\
						 0&-\frac12 & 0 & \frac{\sqrt{3}}{2}
								\\
						0&0&-\frac{ \sqrt{ 3}}{ 2} &0
				\end{pmatrix}\,,\,\,
		 U=\begin{pmatrix}
					\cos\theta &0&0&-\sin\theta
					\\
					0&1&0&0
					\\
					0&0&1&0
					\\
					\sin\theta&0&0&\cos\theta
		 \end{pmatrix}\,.
\end{equation}
Eliminating $M_W$ by using 
\begin{equation}
		 \label{hatA}
	A=\e^{-2M_WM_{\rm N}y} U^\dag \hat A	 
\end{equation}
in a similar way of the analysis done for  the charged gauge boson,
the above EOM becomes
\begin{equation}
	 \label{diagEOM} 
	-m^2 \hat A = \partial_y^2\hat A 
	+\pi RM_G^2 [\delta(y)+\delta(y-\pi R)  ] 
	U\e^{2M_WM_{\rm N}y}U^\dag{\rm diag}(1,0,0,0) U\e^{-2M_WM_{\rm N}y}U^\dag\hat A
	\,.
\end{equation}
It is useful to expand the phase matrix in the following form:
\begin{align*}
	 \e^{-2M_W M_{\rm N} y}
	 =
	 \begin{pmatrix}
		1&0&0&0
		\\
		0&\frac34+\frac14 \cos2M_Wy & -\frac12\sin 2M_Wy & \frac{\sqrt3}{4}(1-\cos2M_Wy)
		\\
		0&\frac12\sin 2M_Wy & \cos 2M_Wy&-\frac{\sqrt3}{2}\sin2M_Wy
		\\
		0&\frac{\sqrt3}{4}(1-\cos2M_Wy)&\frac{\sqrt3}{2} \sin2M_Wy& \frac{1}{4} +\frac{3}{4} \cos2M_Wy  
	 \end{pmatrix}\,.
\end{align*}

Next, we consider the boundary conditions (BCs) at 
$y=0$ and $\pi R$.
We require the {\it $Z_2$ condition at the origin} and {\it periodicity}
similar to those for the charged gauge boson.
Since the $Z_2$ condition on the $\hat A$ are the same as those on the $A$,  
the mode functions $f^b(y)$ satisfy 
\begin{equation}
	 \label{BC_at_origin_1}
	 f(y)={\rm diag}(+,+,-,+)f(-y)\,.
\end{equation}
where $f(y)$ is defined through $\hat A^b_\mu(x^\mu,y)=\hat A^b_{\mu} (x^\mu) f^b(y)$.
By taking into account the condition, 
the EOM (\ref{diagEOM}) are immediately solved in the bulk as  
follows;
\begin{equation}
	f^b(y)
	\propto 
	\begin{cases}
	 \cos (m|y|-\alpha_b) & \text{for } b=0,3,8 \\
	 \sin (my)& \text{for } b=7
	\end{cases}
\end{equation}
where $\alpha_b$ stand for the phases.

Since the delta functions are present at $y=0$ and $\pi R$,
the first derivative of the mode functions becomes discontinuous.
The conditions for the discontinuity are derived by integrating out the EOM (\ref{diagEOM}) around $y=0$ and $\pi R$.
Taking into account the continuous condition at the origin
$\lim_{\epsilon\to 0}\left[A(x,\epsilon)-A(x,-\epsilon)\right]=0$,
we have 
\begin{equation}
	 0
	 =
	 \lim_{\varepsilon\to 0}[\partial_y f^a(x,y)]_{-\varepsilon}^{\varepsilon}   
	 +\pi RM_G^2 f^a \delta_{a0}   
	 \,.
\end{equation}
Note that the index $a$ in the second term does not mean the summation. 
Summarizing the solutions, we find
\begin{equation}
	 \begin{cases}
		f^0\propto \cos(m|y|-\alpha)
		\\
		f^3\propto \cos my
		\\
		f^7\propto \sin my
		\\
		f^8\propto \cos my 
	 \end{cases}
	\,,\, 2\hat m\sin\alpha+\hat M_G^2\cos\alpha=0 
\end{equation}
where $\hat m =\pi R m$ and $\hat M_G=\pi R M_G$\,.  
Same procedure further applies to $y=\pi R$ case.
We integrate out eq (\ref{diagEOM}) around $\pi R-\varepsilon <y< \pi R+\varepsilon$,
it becomes
\begin{align}
		 0=&
		 \lim_{\epsilon\to 0}[\partial_y A]_{\pi R-\varepsilon}^{\pi R +\varepsilon}
		 +\pi RM_G^2U^\dag\text{diag}(1,0,0,0)UA(x,\pi R)   
		\nt \\
		=&
%		-[\partial_y A]_{-\pi R}^{\pi R}   
%		 +\pi RM_G^2U^\dag\text{diag}(1,0,0,0)UA(x,\pi R)   
%		 \\
%		 =&
%		 -[-2M_WM_N\underbrace{\e^{-2M_WM_Ny}U^\dag  \hat A}_{A} ]_{-\pi R}^{\pi R}   
%		-[\e^{-2M_WM_Ny} U^\dag  \partial_y\hat A]_{-\pi R}^{\pi R}   
%		 +\pi RM_G^2U^\dag\text{diag}(1,0,0,0)UA(x,\pi R)   
%		 \\
%		 =&
		-2\e^{-2M_WM_{\rm N}y}U^\dag  \partial_y\hat A(x,y) |_{y=\pi R}^{\text{odd}} 
		 +\pi RM_G^2U^\dag\text{diag}(1,0,0,0)UA(x,y)|^{\text{even} }_{y=\pi R}\,.
%		 \\
%		 \Rightarrow
%		 0=&
%		 \pi R\e^{-2M_WM_Ny}U^\dag  \partial_y\hat A(x,y) |_{y=\pi R}^{\text{odd}} 
%		 -\frac12 \hat M_G^2U^\dag\text{diag}(1,0,0,0)UA(x,y)|^{\text{even} }_{y=\pi R}
\end{align}
The first (second) term stands for  putting $y=\pi R$ after extracting the odd (even) function.
% and {\it vice versa}.
%It is clear that the third  components are automatically satisfied by  respecting parities of the  mode function.
Therefore we have following three relations.
\begin{align}
		 \label{masscondition2}
				0=&
		\Big[
		-\cos\theta\hat m\sin(\hat m-\alpha) -\frac12 \hat M_G^2\cos\theta(1-\frac{3}{4} \sin^2\theta(1-\cos2\hat M_W) )\cos(\hat m-\alpha),
				 \nt\\
				 &~
				 \frac{\sqrt3}{8}\hat M_G^2\sin\theta\cos\theta(1-\cos2\hat M_W)\cos\hat m,
				 \frac{\sqrt3}{4}\hat M_G^2\sin\theta\cos\theta\sin2\hat M_W\sin\hat m, 
				 \nt\\
				 &~
				 -\sin\theta\hat m\sin\hat m-\frac{3}{8}\hat M_G^2\sin\theta\cos^2\theta(1-\cos2\hat M_W)\cos\hat m  
		\Big]  
		\hat A(x) \,, 
		\\
		 \label{masscondition3}
	0=&
		 \Big[
					\frac{\sqrt3}{4}\sin\theta (1-\cos2\hat M_W)\sin(\hat m-\alpha),
					-(\frac34+\frac14\cos2\hat M_W)\sin\hat m,
					\nt\\
					&~
					-\frac12 \sin2\hat M_W \cos\hat m,
					-\frac{\sqrt3}{4}\cos\theta(1-\cos2\hat M_W)\sin\hat m  
		 \Big] \hat A(x) \,,
		 \\
		 \label{masscondition4}
		 0=&
		 \Big[
					\sin\theta(\frac{1}{4}+\frac{3}{4}\cos2\hat M_W)\hat m\sin(\hat m-\alpha)
					+\frac12 \hat M_G^2\sin\theta(1-\frac{3}{4} \sin^2\theta(1-\cos2\hat M_W) )\cos(\hat m-\alpha),
					\nt\\
					&~
					-\frac{\sqrt3}{4}(1-\cos2\hat M_W)\hat m\sin\hat m
					-\frac{\sqrt3}{8}\hat M_G^2\sin^2\theta(1-\cos2\hat M_W)\cos\hat m,
					\nt\\
					&~
					\frac{\sqrt3}{2}\sin2\hat M_W\hat m \cos\hat m
					-\frac{\sqrt3}{4}\hat M_G^2\sin^2\theta\sin2\hat M_W\sin\hat m, 
					\nt\\
					&~
					-\cos\theta(\frac{1}{4}+\frac{3}{4}\cos2\hat M_W)\hat m\sin\hat m
					+\frac{3}{8}\hat M_G^2\sin^2\theta\cos\theta(1-\cos2\hat M_W)\cos\hat m  
		 \Big] \hat A(x)\,.
\end{align}

Next we consider the BCs at $y=\pi R$. The mode functions are continuous at $y=\pi R$ due to the fact that the theory is invariant under the translation along  the extra dimension, namely, 
\begin{align}
	 0=&
		 \lim_{\epsilon\to 0}
		 \left[A(x,\pi R+\epsilon)-A(x,\pi R-\epsilon)\right]
	 =
		 \lim_{\epsilon\to 0}
		 \left[A(x,-\pi R+\epsilon)-A(x,\pi R-\epsilon)\right]
	 \\
	 =&
	 -A(x,y)|^{\text{odd} }_{y=\pi R}   
	  \\
	 =&
	 \begin{pmatrix}
		0\\0\\
		\frac12\sin2\hat M_W\cos \hat m \hat A^{3}+\cos2\hat M_W\sin\hat m\hat A^7
		-\frac{\sqrt3}{2}\sin2\hat M_W (\cos\theta \cos\hat m\hat A^8-\sin\theta\cos(\hat m-\alpha)\hat A^0 )  
		\\0
	 \end{pmatrix}.
\end{align}
In the first line, $\pi R+\epsilon$ is replaced with $-\pi R+\epsilon$
by respecting the translational invariance. 
Expressing by the matrix form, 
we have
\begin{equation}
		 \label{masscondition1}
	0=
	[\frac{\sqrt3}{2}\sin2\hat M_W\sin\theta\cos(\hat m-\alpha),
	 \frac12\sin2\hat M_W\cos\hat m,
     \cos2\hat M_W\sin\hat m,
     -\frac{\sqrt3}{2}\cos\theta\sin2\hat M_W\cos\hat m] 
	 \hat A\,.
\end{equation}

To summarize, the 
(\ref{masscondition2})-(\ref{masscondition4}) and 
(\ref{masscondition1})
give the mass eigenstate and its mass spectrum. 
We subtract them as $(\ref{masscondition4})+\tan(\theta)\times(\ref{masscondition2})+\sqrt{3}\times (\ref{masscondition3}) $ for simplicity.
Replacing $\hat M_G^2 $ with $-2\hat m \tan\alpha$, we have
\begin{equation}
		 0=[0,-\sqrt{3}\sin\hat m , 0 ,-\frac{1}{\cos\theta}\sin\hat m] \hat A
		 .
\end{equation}
By multiplying $\cos\alpha$ by condition (\ref{masscondition2}),
we summarize the conditions 
(\ref{masscondition2})-(\ref{masscondition4}) and 
(\ref{masscondition1})
as
\begin{align}
		 \label{massconditionimp1}
				0=&
		\Big[\frac{\sqrt3}{2}\sin2\hat M_W\sin\theta\cos(\hat m-\alpha),
		 \frac12\sin2\hat M_W\cos\hat m,
     \cos2\hat M_W\sin\hat m,
     -\frac{\sqrt3}{2}\cos\theta\sin2\hat M_W\cos\hat m
		 \Big ] 
		 \hat A \,,
		 \\
		 \label{massconditionimp2}
				0=&
		\Big[
		-\cos\theta\sin(\hat m-2\alpha) -\frac34 \cos\theta\sin^2\theta\sin\alpha(1-\cos2\hat M_W) \cos(\hat m-\alpha),
				 \nt\\
				 &~
				 -\frac{\sqrt3}{4}\sin\alpha\sin\theta\cos\theta(1-\cos2\hat M_W)\cos\hat m,
				 -\frac{\sqrt3}{2}\sin\alpha\sin\theta\cos\theta\sin2\hat M_W\sin\hat m, 
				 \nt\\
				 &~
				 -\sin\theta\cos\alpha\sin\hat m+\frac{3}{4}\sin\alpha\sin\theta\cos^2\theta(1-\cos2\hat M_W)\cos\hat m  
		\Big]  
		\hat A \,, 
		\\
		 \label{massconditionimp3}
	0=&
		 \Big[
					\frac{\sqrt3}{4}\sin\theta (1-\cos2\hat M_W)\sin(\hat m-\alpha),
					\nt\\
					&~
					-(\frac34+\frac14\cos2\hat M_W)\sin\hat m,
					-\frac12 \sin2\hat M_W \cos\hat m,
					-\frac{\sqrt3}{4}\cos\theta(1-\cos2\hat M_W)\sin\hat m  
		 \Big] \hat A \,,
		 \\
		 \label{massconditionimp4}
		 0=&
		 \Big[
					0,
					-\sqrt3 \sin\hat m,
					0,
					-\frac{1}{\cos\theta}\sin\hat m
		 \Big] \hat A 
\end{align}
where $\tan\alpha=-\frac{\hat M_G^2}{2\hat m} $.
Since the brane mass term for the gauge boson $\hat M_G$ is taken to be infinity($\alpha \to -\pi/2$), 
the above conditions become 
\begin{align}
		 \label{massconditionimp01}
				0=&
		\Big[-\frac{\sqrt3}{2}\sin2\hat M_W\sin\theta\sin\hat m,
		 \frac12\sin2\hat M_W\cos\hat m,
     \cos2\hat M_W\sin\hat m,
     -\frac{\sqrt3}{2}\cos\theta\sin2\hat M_W\cos\hat m
		 \Big ] 
		 \hat A \,,
		 \\
		 \label{massconditionimp02}
				0=&
		\Big[
		\sin\hat m-\frac34 \sin^2\theta(1-\cos2\hat M_W) \sin\hat m,
				 \frac{\sqrt3}{4}\sin\theta(1-\cos2\hat M_W)\cos\hat m,
				 \nt\\
				 &~
				 \frac{\sqrt3}{2}\sin\theta\sin2\hat M_W\sin\hat m, 
				 -\frac{3}{4}\sin\theta\cos\theta(1-\cos2\hat M_W)\cos\hat m  
		\Big]  
		\hat A  \,,
		\\
		 \label{massconditionimp03}
	0=&
		 \Big[
					\frac{\sqrt3}{4}\sin\theta (1-\cos2\hat M_W)\cos\hat m,
					-(\frac34+\frac14\cos2\hat M_W)\sin\hat m,
					\nt\\
					&~
					-\frac12 \sin2\hat M_W \cos\hat m,
					-\frac{\sqrt3}{4}\cos\theta(1-\cos2\hat M_W)\sin\hat m  
		 \Big] \hat A \,,
		 \\
		 \label{massconditionimp04}
		 0=&
		 \Big[
					0,
					-\sqrt3 \sin\hat m,
					0,
					-\frac{1}{\cos\theta}\sin\hat m
		 \Big] \hat A\,.
\end{align}
By adopting matrix notation,  they become 
\begin{equation}
		 \label{masscondition_neutral1}
		N \hat A =0 
\end{equation}
where the matrix $N$ is defined by 
\begin{align}
		 \label{masscondition_neutral2}
		N
		=&
		\left[
				 \begin{matrix}
					-\frac{\sqrt3}{2}\sin2\hat M_W\sin\theta\sin\hat m
					&
					\frac12\sin2\hat M_W\cos\hat m
					\\
					\sin\hat m-\frac34 \sin^2\theta(1-\cos2\hat M_W) \sin\hat m
					&
					\frac{\sqrt3}{4}\sin\theta(1-\cos2\hat M_W)\cos\hat m
					\\
					\frac{\sqrt3}{4}\sin\theta (1-\cos2\hat M_W)\cos\hat m
					&
					-(\frac34+\frac14\cos2\hat M_W)\sin\hat m
					\\
					0
					&
					-\sqrt3 \sin\hat m
				 \end{matrix}
		\right.
		\nt \\
		&~~~~~~~~~~~~~~~~~~~~~
		\left.
				 \begin{matrix}
					\cos2\hat M_W\sin\hat m
					&
					-\frac{\sqrt3}{2}\cos\theta\sin2\hat M_W\cos\hat m
					\\
					\frac{\sqrt3}{2}\sin\theta\sin2\hat M_W\sin\hat m
					&
					-\frac{3}{4}\sin\theta\cos\theta(1-\cos2\hat M_W)\cos\hat m  
					\\
					-\frac12 \sin2\hat M_W \cos\hat m
					&
					-\frac{\sqrt3}{4}\cos\theta(1-\cos2\hat M_W)\sin\hat m  
					\\
					0
					&
					-\frac{1}{\cos\theta}\sin\hat m
				 \end{matrix}
		\right]\,.
\end{align}

\subsubsection{mass spectrum}
To have the non-trivial solutions of 
eq(\ref{masscondition_neutral1}),
the mass eigenvalues are obtained from 
solving
\begin{equation}
		 \det N
		 =0.
\end{equation}
Then, we find solutions
\begin{align}
		 \label{spectrum_neutral}
		 \sin^2\hat m =& 0\,,\, 
		 \tan\hat m =
%		 \pm \frac{\sqrt{(3\sin^2\theta-4)(1-\cos2\hat M_W)(3\sin^2\theta\cos2\hat M_W-4\cos2\hat M_W-3\sin^2\theta-4)}}
%{3\sin^2\theta\cos2\hat M_W-4\cos2\hat M_W-3\sin^2\theta}
%\\
%=&
\pm\frac{\sqrt{4\cos^2\theta_W-\sin^2\hat M_W}}{2\cos^2\theta_W-\sin^2\hat M_W}\sin\hat M_W
\,.
\end{align}
We note that the $\theta$ is replaced  with $\theta_W$.
The above result tells us that the neutral gauge bosons split to 
$\gamma, Z'$ (first relation) and $Z$ boson (second relation)
since the right hand side  of the second relation reduce to 
$\frac{\hat M_W}{\cos\theta_W}$
in the limit where the nonlinearity of $v$ can be neglected.

\subsubsection{mass eigenstate}
To find out the mass eigenstate,
we  substitute the corresponding mass eigenvalue
(\ref{spectrum_neutral}) 
for  the conditions (\ref{masscondition_neutral1}).

For  the photon $\gamma$ and anomalous gauge boson $Z'$,
substituting $\sin \hat m=0$ with matrix $N$
leads to 
\begin{equation}
		 \begin{pmatrix}
					0&\frac12&0&-\frac{\sqrt3}{2}\cos\theta\\
					0&\frac12&0&-\frac{\sqrt3}{2}\cos\theta\\
					\frac{\sqrt3}{4}\sin\theta(1-\cos2\hat M_W)&0&-\frac12 \sin2\hat M_W&0
		 \end{pmatrix}
		 \hat A
		 =0
\end{equation}
It gives two different eigenstates. 
They are included as 
\begin{equation}
		 \label{photon_mixing}
		 \begin{cases}
		 \hat A^3_\mu
		 \supset 
		 \sin\theta_W\gamma_\mu \,, \hat A^8_\mu\supset \cos\theta_W \gamma_\mu
		 \\
		 \hat A^0_\mu
		 \supset 
		 \frac{2\cos\theta_W\cos\hat M_W}{\sqrt{4\cos^2\theta_W-\sin^2\hat M_W}}Z'_\mu
		 \, ,
		 \hat A^7_\mu
		 \supset  
		 \frac{\sqrt{4\cos^2\theta_W-1}}{\sqrt{4\cos^2\theta_W-\sin^2\hat M_W}}\sin \hat M_W Z'_\mu
		 \end{cases}
\end{equation}
where $\theta$ is replaced with $\theta_W$.
They are distinguished by taking the limit $\hat M_W\to 0$,
namely, the anomalous gauge boson $Z'_\mu$ is exactly the same as $\hat A^0_\mu$ in the limit 
due to the absence of the mixings by the VEV $v$.
They are also identified by taking the limit $\cos\theta_W\to 1/2\,(\theta\to 0)$
since the brane-localized gauge boson mass term $M_G$ merely couples to 
$A^0_\mu$.

For the $Z$ boson,
substituting the second relation in the eq(\ref{spectrum_neutral}) 
with (\ref{masscondition_neutral1})
leads to
\begin{equation}
		 \label{Z_mixing}
		 \begin{cases}
					\hat A^0_\mu
					\supset 
					\pm \sqrt{\frac{4\cos^2\theta_W-1}{2(4\cos^2\theta_W-\sin^2\hat M_W)}}\sin \hat M_W Z_\mu
					\,,
					\\
					\hat A^3_\mu
					\supset 
					\frac{1}{\sqrt2}\cos\theta_W Z_\mu\,,
					\\
					\hat A^7_\mu
					\supset  
					\mp\frac{\sqrt2 \cos\theta_W\cos\hat M_W}{\sqrt{4\cos^2\theta_W-\sin^2\hat M_W}} Z_\mu
					\,,\\
					\hat A^8_\mu
					\supset 
					-\frac{1}{\sqrt2}\sin\theta_W Z_\mu\,.
		 \end{cases}
\end{equation}
The normalized mode functions are 
\begin{equation}
		 \begin{cases}
					f^0=
					-\frac{1}{\sqrt{\pi R-\frac{1}{2m}\sin2\pi Rm}}\sin m|y|
					\,,\\
					f^3=
					\frac{1}{\sqrt{\pi R+\frac{1}{2m}\sin2\pi Rm}}\cos my
					\,,\\
					f^7=\frac{1}{\sqrt{\pi R-\frac{1}{2m}\sin2\pi Rm}}\sin my
					\,,\\
					f^8=
					\frac{1}{\sqrt{\pi R+\frac{1}{2m}\sin2\pi Rm}}\cos my
					\,.
		 \end{cases}
\end{equation}
The mass eigenvalues $m$ are understood to be substituted by the corresponding ones.  
We then finally solve the mixings as follows.
\begin{align} 
		 \label{mixing0378}
		 \begin{pmatrix}
					A^0\\A^3\\A^7\\A^8	
		 \end{pmatrix}
		 =&
		 \begin{pmatrix}
					\cos\theta\hat A^0+\sin\theta \hat A^8
					\\
					(\frac34 +\frac14\cos2M_Wy)\hat A^3 -\frac12\sin 2M_Wy \hat A^7+
					\frac{\sqrt3}{4}(1-\cos2M_Wy)(\cos\theta\hat A^8 -\sin\theta\hat A^0)   
					\\
					\frac12\sin2M_Wy\hat A^3 + \cos 2M_Wy\hat A^7-\frac{\sqrt3}{2}\sin2M_Wy(\cos\theta\hat A^8-\sin\theta \hat A^0)  
					\\
					\frac{\sqrt3}{4}(1-\cos2M_Wy)\hat A^3+\frac{\sqrt3}{2}\sin2M_Wy\hat A^7
					+(\frac{1}{4}+\frac{3}{4}\cos2M_Wy)(\cos\theta \hat A^8-\sin\theta\hat A^0)  
		 \end{pmatrix}\,,
\end{align}
where $\cos\theta=\frac{\sin\theta_W}{\sqrt3\cos\theta_W},\sin\theta=\frac{\sqrt{4\cos^2\theta_W-1}}{\sqrt3\cos\theta_W}$.

Finally, we comment on the zero mode $Z$ boson. 
The mass spectrum of the KK mode $Z$ boson are split to $m_Z=\pm M_Z +\frac{n}{R}$ according to the condition 
 shown in the eq(\ref{spectrum_neutral}).
The $M_Z$ is a minimum value of $m_z$ which satisfy the above condition. 
So we distinguish them by noting $Z^{(n\pm)}$ but the zero mode $Z^{(0\pm)}$ is degenerate 
since its mass are $\pm M_Z$, respectively.
We therefore regard the $Z^{(0+)}$ and $Z^{(0-)}$ as the SM $Z$ boson.
Then, zero mode $Z$ boson can be read off as  $Z^{(0\pm)}=\frac{1}{\sqrt2}Z$ to avoid the double counting.

In fact, the $Z^{(0\pm)}$ boson are included as
\begin{equation}
\begin{cases}
\hat A^0\supset  \pm \sqrt{\frac{4\cos^2\theta_W-1}{2(4\cos^2\theta_W-\sin^2\hat M_W)}}\sin \hat M_W  f^0_{\pm M_Z}Z^{(0\pm)}
	=\sqrt{\frac{4\cos^2\theta_W-1}{2(4\cos^2\theta_W-\sin^2\hat M_W)}}\sin \hat M_W f^0_{M_Z}Z^{(0\pm)}
	\\
\hat A^3\supset  \frac{1}{\sqrt2}\cos\theta_W f^3_{\pm M_Z} Z^{(0\pm)}
	=\frac{1}{\sqrt2}\cos\theta_W f^3_{M_Z} Z^{(0\pm)}
	\\
\hat A^7\supset  \mp\frac{\sqrt2 \cos\theta_W\cos\hat M_W}{\sqrt{4\cos^2\theta_W-\sin^2\hat M_W}}f^7_{\pm M_Z} Z^{(0\pm)}
	=- \frac{\sqrt2 \cos\theta_W\cos\hat M_W}{\sqrt{4\cos^2\theta_W-\sin^2\hat M_W}}f^7_{M_Z} Z^{(0\pm)}
	\\
\hat A^8\supset -\frac{1}{\sqrt2}\sin\theta_W f^8_{\pm M_Z} Z^{(0\pm)}
	=-\frac{1}{\sqrt2}\sin\theta_W f^8_{M_Z} Z^{(0\pm)}
		\\
\end{cases}
\end{equation} 
where $f^i_{\pm M_Z}= f^i|_{m=\pm M_Z}$.
Then, the quadratic form becomes  $(\hat A^0)^2+(\hat A^3)^2+(\hat A^7)^2+(\hat A^8)^2=(Z^{(0+)})^2+(Z^{(0-)})^2 = Z^2$
and thus we have $Z^{(0\pm)}=\frac{1}{\sqrt2}Z$.

%}}}

%}}}
%%%%%%%%%% end section  %%%%%%%%%%%%%%%%%%%%%%

%%%%%%%%%%%%%%%%%%%%%%%%%%%%%%% bibliography %%%%%%%%%%%%%%%%%%%%%%%%%%%%%%%

%}}}

\end{document}